\documentclass[iop,apj]{emulateapj}

\shorttitle{So You Want To Be An Astronomer:  Astronomy Reading \& You}
\shortauthors{Cooke et al. 2020}

\usepackage{natbib}
\usepackage{graphicx} 
\usepackage{textcomp}
\usepackage{gensymb}
\usepackage{epsfig}%
\usepackage{xcolor}
\usepackage{verbatim}
\usepackage[outdir=./]{epstopdf}
\usepackage{indentfirst}
\usepackage[colorlinks=true,citecolor=blue,linkcolor=blue]{hyperref}
\usepackage{amssymb}
\usepackage{amsmath}
\usepackage{amsfonts}%
\usepackage{fancyhdr}
\usepackage[utf8]{inputenc} 

\begin{document}
\title{Astronomy Paper Seminar\\Participation Guide \& Reading Walkthrough}
\author{Kevin C. Cooke${^1\dagger}$, J. L. Connelly$^2$, K. M. Jones$^1$, Allison Kirkpatrick$^1$, E.A.C. Mills$^1$, Ian J. M. Crossfield$^1$}
\affil{$^1$Department of Physics \& Astronomy, University of Kansas,
Lawrence, KS 66045, USA\\
$^2$Laboratory for Multiwavelength Astrophysics, School of Physics and Astronomy, \\Rochester Institute of Technology, Rochester, NY 14623, USA}
\email{$^\dagger$ Corresp. Author: kcooke@ku.edu}
\begin{abstract}
Welcome to the wonderful world of scientific inquiry!  On this journey you'll be reading many$\times 10^N$ papers in your discipline.  Therefore, efficiency in digesting and relaying this information is paramount.  In this guide, we'll review how you can participate in your local astronomy seminars.  Participation takes many forms, from contributing a recently discovered article to the discussion of a published paper.  In this guide, we'll begin by providing some suggested introductory activities for beginner scientists.  Then we discuss how to locate papers and assimilate their results.  Finally we conclude with a discussion on paper presentation and note storage.  This guide is intended for an undergraduate and graduate student audience, and we encourage faculty to read and distribute this guide to students.
\end{abstract}
\keywords{Astronomical research, Astronomy databases, Astronomical reference materials}

\section{Introductory Activities}\label{sec:intro}
In astronomy seminars and journal clubs across the world, students, postdocs, and faculty discuss the latest astronomy-related news and results.  This can be a daunting thought for new students, as you're managing classes and may not yet feel comfortable in the research world. Fret not, this guide will walk you through how to efficiently read a published paper and relay that information to your peers in addition to the various options you have to contribute to the astronomy discussion.

Before you jump into the deep end and read full papers (Sections \ref{sec:find} \& \ref{sec:read}), consider some of these introductory activities:
\begin{itemize}
\item{Look at the latest images being featured on NASA's Astronomy Picture of the Day\footnote{apod.nasa.gov} website.  This is a good way to get introduced to astronomical objects and phenomenon you may not be familiar with.}
\item{Read one of the papers that someone else has suggested for the week and prepare a question to ask.  This enables you to focus on one particular aspect of a topic you'd like to hear more about.}
\item{Bring an interesting figure you found and ask local experts to explain it; ``This paper had a cool figure, what does it mean?''  This can include a figure from a textbook as well!  There are many historical results that are worth discussing.}
\item{Bring in a news story or press release and ask for the backstory; ``I saw [astronomy thing] in the news. Should I believe it?''  Many popular science articles provide the excellent exposure needed to keep an esoteric field such as ours consistently in the public eye.  However, they can often abbreviate details that you may want to learn more about.}
\end{itemize}

\subsection{Paper Summaries: Astrobites and AAS Nova}
If you don't have the time or don't feel comfortable reading a paper in full, astronomer-run paper summary websites can provide quick summaries on either a given paper or the state of the field.  Some of the most popular of these websites include Astrobites\footnote{astrobites.org} and AAS Nova\footnote{aasnova.org}.  

Astrobites is a daily-updated website hosting summaries of papers and their results, with the summaries written by graduate students.  Astrobites provides short, focused reviews of papers and problems in the field from a student perspective and is a useful starting location for many undergraduate and graduate students.  They also take guest authors.  If you decide scientific writing may be your career path, this is a potential avenue for writing experience!

AAS Nova is similar to Astrobites, but is hosted directly by the American Astronomical Society (AAS) and may be written by either students or more experienced researchers.  As part of a collaborative partnership with Astrobites, AAS Nova will often repost Astrobites articles of noteworthy papers originally published in AAS peer-reviewed journals.

\section{Where to Look For Papers}\label{sec:find}
To find papers on your own, you'll need to get comfortable with online paper databases.  Each scientific field has its own centers where papers are collected for reference and posterity.  For astronomy, this takes the form of (roughly) two central websites.

\subsection{The SAO/NASA Astrophysics Data System (ADS)}
ADS\footnote{ui.adsabs.harvard.edu} (spoken as `ayy-dee-ess') is the first location that many people go to find specific papers of interest.  That's because ADS has the most powerful and versatile search functions.  For example, you'll be able to filter by author name, search for text strings in the abstract, title, main body, and much more.  

The most common search method is by first author name and year. Let's walk through an example. An accomplished scientist you admire tells you to read \citet{Cooke:2019aa} and you know this person works on galaxies.  The most efficient way to find this paper would be to go to the ADS homepage and type in `{\tt $^\wedge$Cooke 2019}' and hit enter.  The carrot will force ADS to search for this author name in the first author space.  There are additional search shortcuts available on the ADS homepage.

Now you come upon the results page and you realize you face a new concern. Cooke isn't a unique name and you don't know the paper title.  The easiest next step is to look at the left sidebar on the results page and start filtering.  Most papers presented in your seminar will be directly related to astronomy, so clicking the astronomy filter cuts out all those pesky physics papers.  Another worthwhile filter to consider is whether the paper is refereed versus non-refereed.  If you're searching for a formally published scientific work such as this example, filtering by refereed will require all the results to have finished the peer review process and have been published in a journal.  The non-refereed choice would be applicable for any in-progress papers not yet through the entire peer review process or informal papers (`white papers') such as conference proceedings. From here, context clues such as the subject of the paper (galaxies are cool) will lead you to the desired paper to complete this example. 

To read the text of the paper, you can find the PDF by first clicking on the paper title, then clicking the PDF icon on the right hand side of the paper's webpage.  You may notice there are two categories, Publisher and arXiv.  Publisher is the formally published version, with all the formatting and spelling corrected for publication.  This link is directly hosted by the publisher, so if you're off your university's network, you may encounter paywalls.  Never pay for a paper, that's the university's job.   To ensure free access to papers outside of the network dominions of universities, a free alternative is hosted on arXiv.

PRO--TIP:  You can add ADS as a search engine within Google Chrome (or your internet browser of choice).  By going into Chrome settings, then `Search Engine', and scrolling down to `Other Search Engines', you can add a new engine keyword.  Click `add', and then name the search engine `ADS', list the keyword as `ads', and input `{\tt https://ui.adsabs.harvard.edu/search/q=\%s}' in the URL space.  Once this is added, you can type `ads' into your URL bar in Chrome, hit tab, and then anything you type afterward will be automatically sent to ADS to search when you hit the enter key.  Shaving a few seconds off by not having to go to the ADS homepage each time will save you days by the time you retire! (Vaddi S., private comm.)  

\subsection{arXiv}
arXiv (spoken just like the word `archive') is an online archive hosted by Cornell University to ensure free access to science papers for those outside of university networks.  While arXiv hosts many subjects, we will often refer to arXiv as `astro-ph' (spoken with the `p' \& `h' as independent letters) because we spend all of our time in the astronomy section\footnote{arxiv.org/archive/astro-ph}.  The key difference between ADS and arXiv is that anyone can upload papers to arXiv, while ADS is an automated system that collects papers from published sources and matching arXiv versions.  arXiv also has less advanced search features in comparison to ADS.

What arXiv does well is host everything everyone is publishing/releasing and collecting them into daily releases.  You can even sign up for email lists of papers delivered right to your inbox!  By going to the arXiv help page\footnote{arxiv.org/help/subscribe}, you can sign up for daily emails with lists of papers related to specific subfields of astronomy, such as Astrophysics of Galaxies, Earth and Planetary Astrophysics, and more!  So sign up for your relevant subfield and peruse these emails when you're curious about the bleeding edge of your subfield's research.  arXiv also assigns each paper a unique number, used in the paper's URL page, to make it easy for people to share papers without having to perform a full search.

Finally, if the giant walls of text that arXiv daily emails updates provide aren't catching your attention, you can try the arXiver\footnote{arxiver.moonhats.com} webpage.  This is an automated webpage that shows the latest uploads to arXiv with a selection of figures included!  You can also use a search category to look at specific subfields such as `Galaxy Astrophysics'.  If you are attempting to keep up to date with the latest results, this figure-focused service may help.

\subsection{Paper Services Summary}
When you need to look for a paper in general, you can use ADS or arXiv relatively equally effectively to search for something to read, with each providing search options for finding recent papers and looking for text strings so you can find subject relevant papers.  If you were told to look for a specific paper and author, ADS will be the easier way to go.  If you are looking for something new, checking arXiv's recent submissions page provides you  papers submitted within the last five business days.

A final recommendation is to read the comments section below the paper abstract on the paper's arXiv or ADS homepage to see if the paper is `Accepted'. This indicates if the paper completed peer review.  New exciting papers may pre-release a version on arXiv prior to acceptance, and the results may be perfectly valid, but you should try and keep in mind what is officially published and what is not.  While it's preferred for seminar and journal club papers to be accepted to a journal, this should not be a hard rule.  There are many white papers on the state of the field that are important for discussion at any institution, yet not included in refereed journals.   For example, critical white papers worth exploring are \citet{Norman:2009aa}'s work on increasing the number of underrepresented minorities in astronomy, \citet{Levesque:2015aa}'s work regarding the efficacy of the Physics GRE, and \citet{Rasmussen:2019aa} summary of gender equity in astronomy.   At the time of writing this guide, many white papers have been recently written as part of astronomy's `Decadal Survey' investigating the state of astronomy as we enter the 2020s.

\section{Methodology of Reading}\label{sec:read}
Now comes the difficult part, actually reading the paper. What makes this an especially pernicious problem is that this process never ends.  Science will continue to progress in an ever expanding literature base and each person learns differently from the form of text.  One constant you can depend on is a general paper structure that has been standardized to ensure that you can identify where relevant information is located.  The main text of a paper starts with an introduction giving an overview and background of the science question.  Next the paper includes a sample selection and/or data section, which includes the fine details of how the data was taken or the simulation was performed.  Unless you are attempting to directly compare data sets, it's alright to focus on the basics here.  The analysis \& results section comes next, with a commonly combined discussion on what tests the authors ran, what correlations the authors found, what properties the authors measured.  This section will focus on the immediate results of the analysis performed.  The physical interpretation of the results and how they fit into the science question of the paper are discussed at length in a discussion section.  This is where the authors make their argument regarding the importance of their results and should be read with a discerning eye. Finally the paper will wrap up all the important points using a conclusions section.

When you read papers, you will each develop your own reading style and system of prioritizing knowledge.  Some people will read a paper directly through each section and figure in its precise order as written by the author (rarely recommended).  Others will instead read the introduction, results, and conclusions first.  Some readers will prioritize the figures first before getting into the details of the text.  As there is no `right way', this guide is to walk you through the logic of what you absolutely need to get out of a paper, and you can experiment with how to retrieve the information on your own.  Some concepts will overlap with each other, this is intended to illustrate how there are numerous forms of reading a paper.

\subsection{Abstracts}\label{sec:abstract}
Before the main text of the paper there is a short summary paragraph called an `abstract'.  An abstract is meant as a distilled version of the paper that provides information on important highlights such as their goals and findings.  Therefore abstracts give you a testing mechanism to see if a paper is relevant to your interests.  Due to the size limitation of abstracts, much of the context of the work must still be found in the main body of the text.   To fully understand the work done in the paper, you'll need to recover important information from the paper yourself.

\subsection{What is the Mystery/Controversy being Addressed?}
The very first thing that you need to discover is why anyone would care about this paper.  This is answered in the introduction and conclusions sections where a good author will summarize the state of the field and an ongoing mystery or controversy.  Usually the second to last paragraph in the introduction famously includes the key phrase, ``Our work will address this problem by \textellipsis".  While it must be rephrased to be relevant to the paper's topic specifically, you'll usually find a clear idea of the paper's goals near the end of the introduction and the beginning of the conclusions. 

If you feel overwhelmed by the many abstract concepts a paper is throwing at you, remember that physical meaning is the key.  All these researchers are spending time investigating the real universe and trying to determine how physical objects such as galaxies/planets evolve.  

{\bf Before moving on to any of the methods or results, you need to feel comfortable saying, `This paper is investigating [this specific problem] with [these objects]'. }

\subsection{How do They Address This Topic?}
The next level of detail you should learn is how the authors go about addressing the problem.  First, find and read through their sample selection section to understand the nature of the objects they are working with.  For example, a paper on spiral star-forming galaxies will be looking at a different population from red spiral galaxies or dwarf spiral galaxies.  As you gain experience in the field, you'll understand how each type of subsample compares to another.  

Next, what observations and derived characteristics of their sample are being used? This is a section where many new students will focus on the arcane details such as exposure time or the specific chemical composition of the detectors for fear of faculty quizzing them about these things.  Focus on the basics:  photometry or spectroscopy, what wavelength regime did they observe, why those wavelengths?  You can always circle back to the details if needed.  Your goal is to understand what the physical meaning behind the measurement is.  Many papers will take some knowledge on the part of the audience for granted, so feel free to ask your advisor questions; `Why did they choose the H$\alpha$ line?', `What components of an object are emitting in the near-infrared?'.  Remember the faculty are there to help you more than you are here to help them.  

{\bf After this stage, you should be able to say, ``To address this problem, they observed [their sample] using [x telescope] so they could determine [physical measurement], which will answer their problem''.}

\subsection{What did they find/not find?}
Finally, you should understand the key results from the paper. These are discussed in depth in the discussion section of a paper, as well as in a condensed form in the conclusions.  Therefore if you're getting lost in the discussion, use the conclusions section as a guide to what the authors want to highlight the most, and then circle back to the detailed discussion if needed.    Usually, these key results will be interpreted from the data found in their plots.  If you find a paper talking at length about one plot in particular, it's bound to be important to their results.  Not all papers break the mold and find something new though!  Perhaps a trend is highly debated and a paper doesn't find it, that's a result indicating that the objects may not be evolving the way we previously thought.  There are many forms to scientific progress.

{\bf By the end of reading these last sections, you should feel comfortable saying, ``The authors find that [these objects] do the following behavior \textellipsis  This is a new result because we previously thought\textellipsis". } 

\section{How to Communicate This Paper to Your Illustrious Colleagues}
Talking about a paper in journal club can be a very anxious experience for many developing scientists because they are concerned about being judged by their peers and more experienced scientists.  What if I miss an important detail?  What if I don't understand something basic?  What if its a bad paper and I can't tell because I don't know what I don't know? Remember that you're a student and it's ok to not know things.  Not only is the process of reading papers meant to educate you, the process of presenting papers is meant to connect you to external expertise.  The journey of a 1000 miles begins with a single step. This principle can also be summarized as: ``Sucking at something is the first step towards being sort of good at something.''

When it's your turn to present your paper, first let people see the title and authorship of your paper so they can look it up on their own devices if they want.  Papers are most commonly identified by first author name and year of publication so get into the habit of identifying papers in this manner.  Then your explanation should include the following highlights:

\begin{itemize}
\item{Explain the physical meaning of the problem being studied, i.e. the big picture: ``This paper is investigating [this specific problem] with [these objects]'.}
\item{Explain the data and analysis they performed: ``To address this problem, they estimated [these physical quantities] by measuring [a certain wavelength range of light, spectra, images, etc.].}
\item{Explain the physical meaning of the results: ``With these measurements, they found [these objects] were exhibiting [such and such a behavior].}
\item{Explain again how this fits into the larger picture: ``This means that these objects are evolving [in a specific way related to the original problem stated in the introduction].}
\end{itemize}

The key is to focus on the physical meaning of the investigation in the paper and don't be afraid to circle back to the original purpose of the experiment so the audience knows why this is important.  You will develop your own style of presenting a paper; this is meant as a rough scaffolding from which you can develop your own.  Don't be afraid to go through the paper in a non-linear fashion, and while you're not expected to know every detail of the paper  (or relay it to the audience), you should always know where to go to find the answer of a question from the audience. 

\subsection{Plots}
Finally, deserving a paragraph in this guide for itself: PLOTS.  The quality and clarity of plots, also called figures, is highly variable, and it's your job to point the audience toward the result the authors found from the same image.  Therefore, go through important plots slowly.  First, review the X and Y axes of the plot with the audience and mention their quantities.  You can't expect the audience to understand the plot if they're reading the axes while you're explaining the image!  After the axes are described, then go through the plot legend and point out which color, line, and/or symbol is the sample of this paper in particular.  Plots will normally include many other literature comparisons.  These are useful to review, but the dataset of this paper should take precedence in your explanation.  Finally, go into the scientific meaning of the plot.  You'll notice this is a lot of setup for what may be a short discussion, but it's strongly recommended to take time describing the plot's axes and legend system for the audience.  Remember that you are not the only student in the room, and there are many subfields in astronomy.  For example, many y-axis labels use subfield specific notations that may make an extragalactic quantity incomprehensible to an exoplanets expert, and vice versa.  Not all scientists are immediately familiar with other subfield's work.

\section{Paper Management}
One thing is certain--as you grow as a scientist, you will have to sort through and manage thousands of papers. You will have papers you find personally interesting, papers that demonstrate a particular methodology, papers that highlight a particular debate, and papers that you use as the basis for your own research. The organizational method that works best for you will be as individualized as how you read the paper in the first place. There are several resources available to help with this.  In this guide, we highlight a few free services available on your personal machine or in the cloud.  These tools will eventually help you write your own papers.  It is key as a scientist to thoroughly and correctly reference related literature.

\subsection{Citation Manager: BibDesk}
BibDesk\footnote{bibdesk.sourceforge.io} is a free citation management software package where you can save the bibliographical information of papers in a single file.  This file is a .bib file directly used to create citations when you write your own works using the standard science paper typesetting system LaTeX (commonly pronounced `lay-tek' and requires a walkthrough paper of its own!).  BibDesk receives citation information by the user directly copy and pasting the citation information from the paper's ADS page, or though BibDesk's web import capability.  BibDesk can instantly copy the citation information of a paper after loading the URL of the paper's ADS homepage.  

BibDesk is a recommended manager not only for its citation management, but through its search function.  You can search the titles for keywords of interest, and BibDesk will return the citations that match your search with a relevance rating.  This can be further improved by adding notes to the keywords field of a paper's bib entry.

Finally, you can create Smart Groups or Static Groups.  These are sub-collections of papers that satisfy a given text search criterion (Smart Group) or a group of papers you manually collect (Static Group).  These groups are useful tools for differentiating between papers you saved due to results, methodology, or were just cool!

\subsection{PDF Managers: Mendeley \& Zotero}
Mendeley\footnote{mendeley.com} \& Zotero\footnote{zotero.org} are paper managers similar to BibDesk, however they have the added benefit of including the full abstract in the paper bib object.   You can manage a folder on your personal machine with the paper .pdf files, and quickly retrieve the pdfs rather than going online again.  Both also have an export function to create a .bib file for use in paper writing.  Note that you have to re-export the .bib file after any changes.  While we limit our recommendations in this guide to free software, check with your department to see if they have purchased an institutional license to an online service that's available to use for students to use for free.  

\subsection{Basic Text Manager: Evernote}
While BibDesk and Mendeley are useful for large collections, perhaps you want to save very detailed information or figures from a small collection of papers.  In this case, a text-based note-keeping service such as Evernote\footnote{evernote.com} may be useful. Not only can you manage text in Evernote, but you can embed images within your notes.  This is useful for recording methodology details or starting a collection of results comparisons before you start writing your own paper.  Due to being giant unformatted text files, Evernote files aren't meant as citation storage locations.  

\section{Concluding Remarks}
We hope that this guide serves as a foundation as you begin to explore the world of astronomy and participate in your local discussions.   We have discussed how to discover and understand astronomy papers, how to store that information, as also how to relay it to others.  Remember, when you have read a paper and are presenting it, you are the local expert on that paper. When people ask questions, it is because they trust your expertise and are looking to you for the answer, not because they are trying to make you look foolish.  You are also not responsible for the content of the paper.  If a colleague has a disagreement with the work presented, do not interpret that as a criticism of you.  

Just as the literature is a diverse realm of refereed papers, white-papers, state-of-the-field overviews and more, you too will develop your own unique way of interpretation.  While it is easy to get lost in abstract concepts, always try to ground yourself in the physical processes being discussed.  It is using these processes that scientists build a connection to our surrounding universe.   Finally, remember to be kind to yourself as you grow as a scientist and learn your newest language, science.

\section{Acknowledgements}
We would like to acknowledge the contribution of the faculty, graduate students, and undergraduate students at the University of Kansas for contributing their opinions and personal experience toward the creation of this document.  This research has made use of NASA's Astrophysics Data System.
\bibliography{CookeKartaltepe2019bib}

\end{document}